УДК 523.92+523.942

# КРУПНОМАСШТАБНЫЕ СВОЙСТВА ТИЛЬТА ГРУПП ПЯТЕН И ЗАКОН ДЖОЯ ВБЛИЗИ СОЛНЕЧНОГО ЭКВАТОРА

К.М. КУЗАНЯН (1,2), Н.Т. САФИУЛЛИН (1,3), Н.И. КЛИОРИН (4,5),
И. РОГАЧЕВСКИЙ (4,5), С.В. ПОРШНЕВ (3)

1) *Крымская астрофизическая обсерватория РАН, 298409, Республика Крым, Бахчисарайский р-н., пгт. Научный*

2) *Институт земного магнетизма, ионосферы и распространения радиоволн им. Н.В. Пушкова Российской академии наук (ИЗМИРАН), г. Троицк, Калужское шоссе, д. 4 , 108840, Россия, г. Москва*

3) *Департамент информационных технологий и автоматики Уральского федерального университета, ул. Мира, 19, 620002, Россия, г. Екатеринбург*

4) *Департамент Механики, Факультет инженерных Наук, Университет им. Бен-Гуриона в Негеве, P. O. Box 653, 84105 Беэр-Шева, Израиль*

5) *Nordita, Королевский технологический институт КТН и Стокгольмский университет, Roslagstullsbacken 23, 10691 Stockholm, Швеция*

## АННОТАЦИЯ

Предложен физический механизм образования угла закрутки (тильта) групп солнечных пятен в процессе образования активных областей под фотосферой Солнца. Детально исследуются факторы, связанные с влиянием сил Кориолиса на крупномасштабные течения в супер-граниляционной конвекции в турбулентной среде. На основе расчетов магнитного поля по модели солнечного нелинейного динамо Kleeorin et al. (2016) и Safiullin et al. (2018) сделаны оценки порядков величины данного эффекта и оценен тильт-угл в диапазоне широт «королевской зоны» пятенной активности. Эта динамо модель основана на балансе мелко-масштабной и крупно-масштабной магнитной спиральностей, которая описывает в деталях процесс образования солнечных пятен за последние пять циклов активности (начиная с 1964 г.), адаптированная для более широкого класса магнитных проявлений солнечной активности. Построены зависимости среднего тильта за эти пять циклов активности и широтно-временные диаграммы распределения этого значения, в целом удовлетворяющие закону Джоя (Hale et al. 1919, ApJ, 49, 153), но также и показывающие локальные отклонения от него в ограниченном диапазоне широт в отдельных фазах солнечного цикла.

# LARGE SCALE PROPERTIES OF TILT OF SUNSPOT GROUPS AND JOY'S LAW NEAR THE SOLAR EQUATOR

K.M. KUZANYAN (1,2), N. SAFIULLIN(1,3), N. KLEEORIN(4,5), I. ROGACHEVSKII(4,5), S. PORSHNEV(3)

1) *Crimean Astrophysical Observatory of Russian Academy of Sciences, Republic of Crimea, Bakhchisaray District, settl. Nauchniy*

2) *Pushkov Institute of Terrestrial Magnetism, Ionosphere and Radio Wave Propagation, Russian Academy of Sciences, Russia, 108840, Moscow, Troitsk, Kaluzhskoe Hwy 4, IZMIRAN*

3) *Department of Information Technology and Automation, Ural Federal University, 19 Mira str., 620002 Ekaterinburg, Russia*

4) *Department of Mechanical Engineering, Ben-Gurion University of the Negev, P. O. Box 653, 84105 Beer-Sheva, Israel*

5) *Nordita, KTH Royal Institute of Technology and Stockholm University, Roslagstullsbacken 23, 10691 Stockholm, Sweden*

## ABSTRACT

We present a physical mechanism of formation of tilt angles of sunspots due to the process of formation of active regions below the solar photosphere. The contribution of Coriolis force factors on large-scale flows of super-granular convection in turbulent media has been investigated in details. On the basis of earlier works by Kleeorin et al. (2016) and Safiullin (2018) we give physical estimates of orders of magnitude of the effect and estimate the tilt angles near the solar equator, in the "Royal" zone of solar activity. The above model is based on the balance of the small-scale and large scale magnetic helicities and describes in details the sunspot formation process over the last five solar cycles (since 1964). We adopt this model for a wider class of manifestations of solar activity. We present latitudinal dependence of the mean tilt on these five solar cycles and time-latitudinal diagrams over a limited range of latitudes and phases of the solar cycle.

# 1. ВВЕДЕНИЕ

К числу основных закономерностей, которым подчиняются солнечные пятна, относится закон Джоя для пар ведущего и ведомого пятен. Этот закон проявляется в том, что угол наклона этих пар относительно солнечного экватора (тильт) имеет характерное различие по полушариям: ведущие пятна в обоих полушариях находятся, как правило, ближе к экватору, чем ведомые, что означает изменение знака тильта. Тильт групп солнечных пятен является их важнейшей характеристикой, которая используется в построении феноменологических моделей солнечной цикличности, типа модели Бабкока-Лейтона (Babcock, 1961; Leighton 1969). Хотя со времени открытия правила Джоя прошло около ста лет, до сих пор не существует однозначного объяснения этой закономерности.

Простейшее объяснение правила Джоя может быть связано с прямым воздействием сил Кориолиса на движущиеся магнитные трубки. На всех широтах, за исключением экваториальной зоны, это воздействие определяется параметром Кориолиса, который пропорционален синусу гелио-широты. Таким образом, в первом приближении можно ожидать, что закон Джоя также будет пропорционален синусу гелио-широты. Тем не менее наблюдаются более сложные свойства крупномасштабного распределения тильта, которые можно рассматривать как отклонения закона Джоя. В данной работе мы показываем, что эти отклонения связаны с влиянием магнитных сил (сила Ампера), порождаемых крупномасштабным, главным образом тороидальным, магнитным полем. С одной стороны они требуют уточнения данной теории, а с другой стороны, сила Кориолиса не входит непосредственно в уравнение магнитной индукции, поэтому остается только удивляться, почему закон Джоя имеет в первом приближении такую простую синусо-подобную форму. Теория правила Джоя была выдвинута, например, в статье D'Silva & Choudhuri (1993, A&A, 272,621) и развивалась в ряде последовавших работ. Важной особенностью этого цикла работ является использование механизма неустойчивости Кельвина-Гельмгольца для формирования солнечных пятен, а также гипотеза о существовании в глубинах конвективной зоны Солнца сгустков сильного магнитного поля с напряженностями порядка сотен кГс.

Тем не менее, эта картина не согласуется с современными знаниями о Солнце и требует пересмотра. Прежде всего, согласно данным гелиосейсмологии, полученным с помощью космических аппаратов SOHO и SDO за последние 20 лет (см. обзор Косовичева и др. 2000) на Солнце вряд ли возможно существование таких сильных полей. Недавно Сингом и Бранденбургом (2018) было установлено, что процессы формирования активных областей на Солнце разворачиваются в непосредственной близости от поверхности Солнца, на глубинах ~20Мм, едва превосходящих глубину залегания супер-гранул, что подтвердило ранние результаты Косовичева и др. (2000). Это открывает возможность построения моделей пятен с локализованным закручиванием в процессе их эволюции и всплытия.

Кроме того, наблюдения магнитных полей на поверхности Солнца, т. е. магнитных полей солнечных пятен и активных областей, ясно указывают на их спиральную природу. Количественное исследование этой спиральности имеет многолетнюю историю, см., например, Seehafer (1990), Pevtsov & Canfield (1994), Bao & Zhang (1998), Hagino & Sakurai (2004). Восемь лет назад Жанг и др.

(2010, MNRAS) (далее Z10) предприняли систематическое исследование токовой спиральности и закрутки (твиста) фотосферных магнитных полей внутри солнечных активных областей. Полученные в Z10 двухмерные баттерфляй-диаграммы (гелиоширота – время) построены для радиальной составляющей токовой спиральности $B_r \left[ \text{rot}(\mathbf{B}) \right]_r$ и твиста (скрученности) магнитного поля, усреднённых по статистически значимому ансамблю магнитных полей активных областей, покрывающих почти два цикла солнечной активности.

С другой стороны, в текущей литературе имеется значительная коллекция статей, посвященных наблюдениям общей закрутки активных областей (тильту). Эти наблюдения (Stenflo & Kosovichev, 2011) показывают, что с одной стороны закон Джоя выполняется достаточно хорошо: тильт $\delta \propto \sin \varphi$ ($\varphi$ - гелиографическая широта), а с другой - на низких широтах наблюдаются заметные отклонения от этого закона (Тлатова и др. 2015; Tlatova et al. 2018). Наконец, активные области малых размеров демонстрируют анти-джоевское поведение: $\delta \propto -\sin \varphi$ (Тлатов др., 2013; 2015). В данной работе мы попытаемся дать теоретическое объяснение этим странностям в поведении тильта, используя результаты модели нелинейного динамо и элементарные соображения о закручивании полей активных областей силой Кориолиса.

Полученные результаты могут представлять интерес не только для описания свойств закрутки активных областей (тильта), но и возможно позволят оценить поток магнитной спиральности в хромосферу и корону и, тем самым, энергетический потенциал вспышечной активности Солнца.

## 2. МОДЕЛЬ БИПОЛЯРНОЙ АКТИВНОЙ ОБЛАСТИ

В данной работе мы рассмотрим в качестве модели простую биполярную активную область, в которой расстояние между областями противоположной полярности составляет $L$, и существующую в верхней части супергрануляционной ячейки турбулентной конвекции глубиной $L/2$. Последнюю мы будем связывать с высотой равновесной фотосферы по плотности $L/2 \approx H_\rho = -\left[ d \log \rho_0 (r)/dr \right]^{-1}$, где $\rho_0$ - равновесная плотность солнечной плазмы на глубине формирования солнечных пятен $H = 10^9$ см =10 Мм. Согласно Чандрасекхару (1962, Chap. 16, p. 48, Fig. 7a) совокупность трёх систем ролов размером $L/2$ на $L/2$ с осями, повёрнутыми друг к другу под углом $120°$, образуют систему гексагонов, каждый из которых имеет максимальный размер $L$ и глубину $L/2$. Таким образом, каждый гексагон может быть вписан в круг радиуса $L/2$. Следует, однако, иметь в виду, что, во-первых, реальные солнечные и звёздные конвективные ячейки, согласно наблюдениям солнечной поверхности, имеют скорее форму неправильных пентагонов и гексагонов, чем абсолютно регулярных гексагонов, а во-вторых, их горизонтальные размеры, по-видимому, превышают, примерно вдвое, размеры классических ячеек Релея. Причина в том, что оптимальные ролы в сильно турбулентной конвекции имеют в сечении размер не $L/2$ на $L/2$, как в ламинарной конвекции, а приблизительно $L$ на $L/2$ (Эльперин и др. 2000, Букаи и др.2009). Таким образом, реальный размер супер-гранул должен быть порядка $2L$, т.е. быть в 4 раза больше её глубины $H_\rho$ и составлять $\approx 40$ Мм. Данные численные оценки носят, безусловно, иллюстративный характер и диапазон этих значений может варьироваться.

Рассмотрим теперь систему уравнений, описывающую поведение (в том числе, закрутку) магнитных полей активной области. Это уравнение движения (уравнение Навье-Стокса), уравнение индукции и уравнение переноса энтропии в неупругом приближении. Согласно Вандакурову (1971), эта система уравнений имеет вид:

$$\frac{\partial \mathbf{V}}{\partial t} = -\vec{\nabla}\left(\frac{\overline{P}}{\rho_0} + \frac{V^2}{2} + \frac{H^2}{8\pi\rho_0}\right) - \mathbf{g}\,S + \mathbf{\Lambda}\frac{H^2}{8\pi\rho_0} + \frac{1}{4\pi\rho_0}(\mathbf{H}\cdot\nabla)\mathbf{H} + \mathbf{V}\times(2\mathbf{\Omega}+\mathbf{W}) \qquad (1.1)$$

$$\frac{\partial \mathbf{H}}{\partial t} = (\mathbf{H}\cdot\nabla)\mathbf{V} - (\mathbf{V}\cdot\nabla)\mathbf{H} - \mathbf{H}(\mathbf{V}\cdot\mathbf{\Lambda}) \qquad (1.2)$$

$$\frac{\partial S}{\partial t} + (\mathbf{V}\cdot\nabla S) = -\frac{\Omega_B^2}{g}V_z \qquad (1.3)$$

$$\operatorname{div}(\mathbf{V}) = \mathbf{\Lambda}\cdot\mathbf{V} \equiv -\frac{\nabla \rho_0}{\rho_0}\cdot\mathbf{V} \qquad (1.4)$$

где $\mathbf{V}, \mathbf{H}, S, \overline{P}$ – гидродинамическая скорость, магнитное поле, энтропия и давление плазмы активной области, соответственно. Здесь первый член в правой части (1.1) – сила полного давления, $-\mathbf{g}\,S$ и $\mathbf{\Lambda}H^2/8\pi\rho_0$ – силы гидродинамической и магнитной плавучести соответственно, $(\mathbf{H}\cdot\nabla)\mathbf{H}/4\pi\rho_0$ – сила магнитных натяжений, $\mathbf{V}\times(2\mathbf{\Omega}+\mathbf{W})$ - обобщённая сила Кориолиса, включающая в себя не только регулярное вращение Солнца (звезды) $\mathbf{\Omega}$, но и локальную завихрённость течения $\mathbf{W}=\operatorname{rot}(\mathbf{V})$, $\Omega_B^2$ – квадрат частоты Брунта-Вяйселля. Напомним, что в конвективных зонах $\Omega_B^2 < 0$, а в фотосфере и выше, где нет конвекции, $\Omega_B^2 > 0$, а все силы, упомянутые здесь - на единицу массы. Поскольку нас интересует поведение этой системы в фотосфере Солнца и выше, где, по-видимому, нет ни магнито-гидродинамической (MHD–) турбулентности, ни динамо, мы пренебрежём магнитной и кинематической вязкостью плазмы и коэффициентом диффузии энтропии. Будем искать решение системы (1.1-4) в виде $\mathbf{V}=\mathbf{u}+\mathbf{v}$, $\mathbf{H}=\mathbf{B}+\mathbf{b}$, $S=\overline{S}+s$, $\overline{P}=P+p$, где $\mathbf{u}, \mathbf{B}, \overline{S}, P$ – решение системы (1.1-4) при $\mathbf{\Omega}=0$. Поскольку тильт мал, (он редко когда превосходит $10°$), то можно считать поля $\mathbf{v}$, $\mathbf{b}, s, p$ малыми возмущениями по сравнению с равновесными полями $\mathbf{u}, \mathbf{B}, \overline{S}, P$. Тогда, вводя новую переменную $\boldsymbol{\xi}$: $\mathbf{v}=\partial\boldsymbol{\xi}/\partial t$, имеющую смысл смещения плазменного элемента относительно равновесного положения, и интегрируя с помощью неё уравнения (1.2-3), получаем следующее элегантное уравнение относительно $\boldsymbol{\xi}$:

$$\frac{\partial^2 \boldsymbol{\xi}}{\partial t^2} = -\vec{\nabla}\left(\frac{p_{tot}}{\rho_0}\right) - \hat{\mathbf{r}}\xi_z\left[\Omega_B'^2 + \Lambda^2 V_A^2\right] + 2\left(\mathbf{u} + \frac{\partial\boldsymbol{\xi}}{\partial t}\right)\times\mathbf{\Omega} + (\mathbf{V}_A\cdot\nabla)^2\boldsymbol{\xi} \qquad (1.5)$$
$$+ \mathbf{\Lambda}\mathbf{V}_A\cdot(\mathbf{V}_A\cdot\nabla)\boldsymbol{\xi}$$

Здесь через Альфвеновскую скорость $\mathbf{V}_A = \mathbf{B}/\sqrt{4\pi\rho_0}$, $\Omega'^2_B = \Omega_B^2 + g(\hat{\boldsymbol{\xi}}\cdot\nabla S)/(\hat{\boldsymbol{\xi}}\cdot\hat{\mathbf{r}})$ включено магнитное поле. Заметим, что $\boldsymbol{\omega} = \mathrm{rot}(\mathbf{v}) = \partial\mathrm{rot}(\boldsymbol{\xi})/\partial t$. Введем вектор закрутки нашей биполярной активной области $\boldsymbol{\delta}$: $\boldsymbol{\omega} = \mathrm{rot}(\mathbf{v}) = \partial\boldsymbol{\delta}/\partial t = \partial\mathrm{rot}(\boldsymbol{\xi})/\partial t$, тогда: $\boldsymbol{\delta} = \mathrm{rot}(\boldsymbol{\xi})$. Тогда, физический смысл вектора $\boldsymbol{\delta}$ следующий: его величина - это малый угол, на который поворачиваются магнитные силовые линии магнитного поля $\mathbf{B}$ за время $\Delta t$ в случае, если завихрённость $\boldsymbol{\omega} = \mathrm{rot}(\mathbf{v})$ не равна нулю. Направление $\boldsymbol{\delta}$ совпадает с направлением $\boldsymbol{\omega}$, т.е. перпендикулярно плоскости поворота. Таким образом, искомый тильт $\delta$ можно отождествить с радиальной компонентой вектора $\boldsymbol{\delta}$ на границе фотосфера - конвективная зона: $\delta = \delta_r(0)$. Вычислим rot от левой и правой части уравнения (1.5) и спроецируем получившееся векторное уравнение на равновесное поле $\mathbf{B}$ т.е. найдём $\delta_B = \boldsymbol{\delta}\cdot\mathbf{B}/B$ Это даёт:

$$\frac{\partial^2 \delta_B}{\partial t^2} = 2\left\{\mathrm{rot}\left[\left(\mathbf{u}+\frac{\partial\boldsymbol{\xi}}{\partial t}\right)\times\boldsymbol{\Omega}\right]\right\}_B + (\mathbf{V}_A\cdot\nabla)^2\delta_B \qquad (1.6)$$

Здесь нижний индекс обозначает проекцию на вектор магнитного поля $\mathbf{B}$. Это уравнение описывает скручивание поля под влиянием силы Кориолиса на конвективные течения. Рассмотрим свободное решение уравнения (1.6) в пренебрежении слагаемым $2\mathrm{rot}[(\partial\boldsymbol{\xi}/\partial t)\times\boldsymbol{\Omega}]$. Его общее решение описывает стоячие альфеновские волны; оно даёт максимальную закрутку вблизи оснований магнитных петель в виде:

$$\delta_B = \sum_{n=0}^{\infty} A_n \cos\left(\frac{\pi(2n+1)}{L_B}\zeta\right)\cos\left(V_A\frac{\pi(2n+1)}{L_B}t + \varphi\right),$$

и описывает незатухающие колебания с периодами $T_n = 2L_B/(2n+1)V_A = 2\tau_D/(2n+1)$. Здесь $\zeta$ координата, отсчитанная вдоль силовой линии длиной $L_B$, $\tau_D = L_B/V_A$. Взяв для оценки следующие параметры $B = 300$ Гс, $L_B \approx 2L = 40$ Мм, плотность плазмы $\rho_0 \approx 4.5\times 10^{-7}$ г/см$^3$ (Baker and Tameswary, 1966) получим $V_A \approx 1.5$ км/с, $\tau_D = (2\div 4)\times 10^4$ с = 6 - 12 часов = 0.25 – 0.5 дня. Таким образом, для времени закручивания активной области получим: $\tau_D = 0.25 - 0.5$ дня $\ll T_{27.3}$ много короче (Кэррингтоновского) периода вращения Солнца вокруг своей оси. Поэтому слагаемое $2\mathrm{rot}[(\partial\boldsymbol{\xi}/\partial t)\times\boldsymbol{\Omega}]$ пренебрежимо мало. Предположим, что на границе конвективной зоны и фотосферы $I(\zeta) = 2\mathrm{rot}_B[\mathbf{u}\times\boldsymbol{\Omega}]$ быстро обращается в ноль вне границы. Тогда источник в левой части уравнения (1.6) - есть приближённо комбинация двух узких импульсов: $I(\zeta) = 2\mathrm{rot}_r[\mathbf{u}\times\boldsymbol{\Omega}][\delta(\zeta) - \delta(-L_B + \zeta)]$, где $\delta(\zeta)$, функция Дирака. Раскладывая источник $I(\zeta)$ в ряд Фурье на отрезке $[0, L_B]$ в базисе $\pi(2n+1)/L_B$ получим: $I_n = 4\mathrm{rot}_r[\mathbf{u}\times\boldsymbol{\Omega}]/\pi \equiv I/\pi$. Подставляя это выражение в (1.6) и раскладывая решение в интеграл Фурье получим для каждой Фурье-составляющей:

$$\frac{\partial^2 A_n}{\partial t^2} = \frac{I}{\pi} - \left[V_A \frac{\pi(2n+1)}{L_B}\right]^2 A_n \qquad (1.7)$$

Общее решение (1.7) с начальным условием $A_n(0) = 0$ имеет вид:

$$A_n = I \frac{\tau_D^2}{\pi^3 (2n+1)^2} \left[1 - \cos\left(\frac{\pi(2n+1)}{\tau_D} t\right)\right].$$

Вычислим $\delta_B$

$$\delta_B = \frac{I \tau_D^2}{\pi^3} \sum_{n=0}^{\infty} \frac{1}{(2n+1)^2} \left[1 - \cos\left(\frac{\pi(2n+1)}{\tau_D} t\right)\right] \cos\left(\frac{\pi(2n+1)}{L_B} \zeta\right).$$

Подсчитаем тильт $\delta = \overline{\delta}_B(0)$ ( линия над $\overline{\delta}_B(0)$ означает усреднение по времени):

$$\overline{\delta}_B(0) = 4 \operatorname{rot}_r [\mathbf{u} \times \mathbf{\Omega}] \frac{\tau_D^2}{\pi^3} \sum_{n=0}^{\infty} (2n+1)^{-2} = 4 \operatorname{rot}_r [\mathbf{u} \times \mathbf{\Omega}] \frac{\tau_D^2}{\pi^3} \frac{\pi^2}{8} = \frac{\tau_D^2}{2\pi} \operatorname{rot}_r [\mathbf{u} \times \mathbf{\Omega}]$$

Вычислим $\operatorname{rot}_r [\mathbf{u} \times \mathbf{\Omega}]$:

$$\operatorname{rot}_r [\mathbf{u} \times \mathbf{\Omega}] \approx (\mathbf{\Omega} \cdot \nabla) u_r - \Omega_r \operatorname{div}(\mathbf{u}) = -\Omega \left[\cos(\theta)\left(\frac{u_r}{H_\rho} - \frac{\partial u_r}{\partial r}\right) - \frac{1}{r} \frac{\partial u_r}{\partial \theta} \sin(\theta)\right]. \quad (1.8)$$

Для оценки поведения скорости в случае супергрануляции часто используют закон непрерывности потока импульса в следующей форме $\partial(r^2 \rho_0 u_r)/\partial r = 0$ (см., например, Гибсон). Это даёт:

$$\frac{\partial u_r}{\partial r} = \frac{u_r}{H_\rho} - \frac{2 u_r}{r} \approx \frac{u_r}{H_\rho}$$

Подобная оценка, безусловно, справедлива у основания супергранулы, но не вблизи поверхности, где знак $\partial u_r / \partial r$ должен поменяться, поскольку скорость $u_r$ должна упасть до нуля. Поэтому, мы примем для оценки:

$$\frac{\partial u_r}{\partial r} \approx -k \frac{u_r}{H_\rho},$$

где $k \sim 1$ Подставляя это в (1.8) получим:

$$\operatorname{rot}_r [\mathbf{u} \times \mathbf{\Omega}] \approx -\Omega(1+k) \tau_F^{-1} \left[\cos(\theta) + \frac{\tau_F}{(1+k) r} \frac{\partial u_r}{\partial \theta} \sin(\theta)\right], \qquad (1.9)$$

где $\tau_F = H_\rho / u_r$. Вертикальная составляющая скорости имеет, по крайней мере, два вклада: регулярный и случайный: $u_r = U_r + u_r^{(c)}$, соответственно. Регулярную составляющую следует связать с солнечной меридиональной циркуляцией, в том числе, зависящей от фазы цикла. Случайная составляющая порождается случайным движением плазмы супергранул. Мы оценим её вклад следующим образом: $\partial u_r^{(c)} / \partial \theta \times \tau_F / r (1+k) \approx \approx u_r^{(c)} / L (1+k) \times H_\rho / u_r^{(c)} = H_\rho / \xi L (1+k) \approx \xi / 2(1+k)$, где $\xi$ - случайная функция порядка единицы с нулевым средним $\langle \xi \rangle = 0$. В результате вместо (1.9) имеем:

$$\operatorname{rot}_r\left[\mathbf{u}\times\boldsymbol{\Omega}\right] \approx -\Omega(1+k)\tau_F^{-1}\left[\cos(\theta)+\left(\frac{\xi}{2(1+k)}+\frac{\tau_F}{R_\odot}\frac{\partial U_r}{\partial \theta}\right)\sin(\theta)\right] \qquad (1.10)$$

Подставляя (1.10) в выражение для тильта, порождаемого вращением, получим:

$$\delta = \frac{\tau_D^2}{2\pi}\operatorname{rot}_r\left[\mathbf{u}\times\boldsymbol{\Omega}\right] \approx -\frac{(1+k)\tau_D^2}{\tau_F T_\odot}\left[\sin(\varphi)+\cos(\varphi)\left(\frac{\xi}{2(1+k)}-\frac{\tau_F}{R_\odot}\frac{\partial U_r}{\partial \varphi}\right)\right], \quad (1.11)$$

Здесь мы перешли от ко-широты $\theta$ к гелиоцентрической широте $\varphi = \pi/2 - \theta$ и $T_\odot$ -период вращения Солнца вокруг своей оси. Оценим коэффициент перед скобкой в (1.11) : он должен быть порядка $1/3$. Течения, порождаемые средним полем, вычислены в работе Клиорина и Рузмайкина (1991). Согласно их вычислениям

$$U_r \approx \frac{\ell_0(z)}{4\pi\rho\nu_T R_\odot}\left(\frac{zH_\rho(z)}{R_\odot}+\frac{\ell_0(z)}{1-2\varepsilon}\frac{R_\odot^2}{r^2}\right)\frac{1}{\sin\theta}\frac{\partial}{\partial\theta}(\sin\theta F(\theta)), \qquad (1.12)$$

где функция $F(\theta)$ имеет вид:

$$F(\theta) = \int_{R_\odot - H_\odot}^{R_\odot}\left[\left(1+\frac{R_\odot - r}{H_\odot - \ell_0(H_\odot)}\right)\left(\frac{1}{r}\frac{\partial}{\partial\theta}-\cot(\theta)\frac{\partial}{\partial r}\right)B_\varphi^2\right]dr.$$

Действуя в рамках «no-r» модели (Соколов и др. 1995) мы можем переписать это выражение следующим образом:

$$F(\theta) = \int_{R_\odot - H_\odot}^{R_\odot}\left[\left(1+\frac{R_\odot - r}{H_\odot - \ell_0(H_\odot)}\right)\left(\frac{1}{r}\frac{\partial}{\partial\theta}+\cot(\theta)\frac{\mu}{R_\odot}\right)B_\varphi^2\right]dr,$$

где $\mu_\odot = R_\odot/H_\odot$, $\ell_0(H_\odot)$, $\ell_0(z)$ - длина пути перемешивания на глубине конвективной зоны $H_\odot$ и на глубине $z$ соответственно. Можно принять, что $\ell_0(H_\odot)\simeq 80\,\mathrm{Mm}$; толщина конвективной зоны $H_\odot\simeq 200\,\mathrm{Mm}$; $R_\odot\simeq 700\,\mathrm{Mm}$; Тогда $R_\odot/(H_\odot - \ell_0(H_\odot))\simeq 5.9$. Согласно (1.11), нас интересуют течения только у поверхности. Его можно переписать в виде:

$$U_r \approx \frac{\ell_0^2(0)}{4\pi\rho\nu_T(1-2\varepsilon)R_\odot}\left[\frac{\partial^2 f(\varphi)}{\partial\varphi^2}-\mu_\odot\tan(\varphi)\frac{\partial f(\varphi)}{\partial\varphi}+\mu_\odot\frac{f(\varphi)}{\cos^2(\varphi)}\right] \qquad (1.13)$$

Напомним, что мы перешли к гелиоцентрической широте $\varphi = \pi/2 - \theta$. Легко убедиться, что

$$f(\varphi) = \int_{R_\odot - H_\odot}^{R_\odot}B_\varphi^2\frac{dr}{r}+\frac{R_\odot}{H_\odot - \ell_0(H_\odot)}\int_{R_\odot - H_\odot}^{R_\odot}\left[\frac{R_\odot - r}{R_\odot}\right]B_\varphi^2\frac{dr}{r}$$

Подставляя (1.13) в (1.12) получим:

$$\delta \approx -\frac{(1+k)\tau_D^2}{\tau_F T_\odot}\left[\sin(\varphi)+\left(\frac{\xi}{2(1+k)}-\frac{\tau_F \ell_0^2(0)\Im(f)}{4\pi\rho\nu_T(1-2\varepsilon)R_\odot^2}\right)\cos(\varphi)\right],$$

где функция имеет размерность квадрата магнитного поля:

$$\Im(f) = \frac{\partial^3 f(\varphi)}{\partial \varphi^3} + \mu_\odot \tan(\varphi)\left(\frac{2f(\varphi)}{\cos^2(\varphi)} - \frac{\partial^2 f(\varphi)}{\partial \varphi^2}\right)$$

Величина $\varepsilon$ связана с граничным условием для скорости $U_r$ (Клиорин и Рузмайкин., 1991):

$$\left[\varepsilon \frac{\partial N}{\partial r} + \frac{N}{r}\right]_{r=R_\odot} = 0, \quad N = \rho(r) U_r(r,\theta) \qquad (1.14)$$

Заметим, что $\varepsilon = 0$ соответствует «жесткому» граничному условию $U_r|_{r=R_\odot} = 0$, $\varepsilon = \infty$ - «плоское мягкое» граничное условие $\frac{\partial N}{\partial r} = 0$. Условие (1.14) можно переписать в виде:

$$\left.\frac{\partial N r^{1/\varepsilon}}{\partial r}\right|_{r=R_\odot} = \left\{\frac{\partial}{\partial r}\left[r^{1/\varepsilon} \rho(r) U_r(r,\theta)\right]\right\}\bigg|_{r=R_\odot} = 0 .$$

При $\varepsilon = 0.5$ получается «сферическое мягкое» граничное условие:

$$\left.\frac{\partial N r^2}{\partial r}\right|_{r=R_\odot} = \left\{\frac{\partial}{\partial r}\left[r^2 \rho(r) U_r(r,\theta)\right]\right\}\bigg|_{r=R_\odot} = 0 .$$

Физически это соответствует тому, что радиальный поток крупномасштабного импульса $N = \rho(r)U_r(r,\theta) = 0$ «не замечает границы» раздела конвективной зоны и фотосферы. Очевидно, что в таком приближении второй член в (1.12) описывает свободный разлёт плазмы Солнца в вакуум в почти гидростатическом неупругом приближении в присутствии стационарного магнитного поля и в почти стационарном, относительно скорости приближении. Этот член может дать только бесконечную стационарную скорость. С учётом ограниченности наших приближений можно быть уверенным, что скорость останется конечной из-за возникновения разрывов и ударных волн при приближении к альфвеновской скорости и скорости звука. Оценим $f(\varphi)$. Несколько простых моделей (например, $B_\varphi \approx B_0(\theta)(R_\odot/r)^{\beta(\theta)}$, $\beta(56°5) \simeq 3.5$, Иванова и Рузмайкин, (1976), $B_\varphi \approx B_0 \cos(\pi(R_\odot - r)/(2H_\odot))$ ), дают $f(\varphi) \approx a_*\left[B_\varphi(R_\odot - H_\odot)\right]^2$, здесь $B_\varphi(R_\odot - H_\odot)$ - поле у основания конвективной зоны. Коэффициент $a_*$ может быть слегка больше единицы, если поле спадает медленнее чем $r^{-1}$, или заметно меньше, если поле спадает быстрее чем $r^{-1}$. Поле $B_\varphi(R_\odot - H_\odot) \simeq B_0 \sim \eta_T/R_\odot \sqrt{4\pi\rho_{R_\odot - H_\odot}/\mu}$ было оценено в работе Клиорина и др. (1995). Здесь параметр $\mu \approx 0.1$ и связан с эволюционным уравнением для мелкомасштабной магнитной спиральности, полученной из закона сохранения полной магнитной спиральности (Клиорин и Рузмайкин., 1982; Клиорин и др., 1995). Будем считать, что турбулентная диффузия и вязкость равны $\eta_T \approx \ell_0^2/3\tau_0$, $\nu_T \approx \ell_0^2/6\tau_0$ (Kl-Ru, Sol. Phys., 1991), соответственно. Записывая

$$f(\varphi) \approx a_*\left[B_\varphi(R_\odot - H_\odot)\right]^2 \approx a_*\left(4\pi\rho_{R_\odot - H_\odot}/\mu\right)\left(\eta_T/R_\odot\right)^2 b_\varphi^2(\varphi) ,$$

и подставляя это в выражения для тильта получим:

$$\delta \approx -\frac{(1+k)\tau_D^2}{\tau_F T_\odot}\left[\sin(\varphi)+\left(\frac{\xi}{2(1+k)}-\frac{2a_*}{3\mu(1-2\varepsilon)}\frac{\rho_{R_\odot-H_\odot}}{\rho_0}\frac{\tau_F \ell_0^4}{\tau_0 R_\odot^4}\frac{\partial^3\left[b_\varphi^2(\varphi,t)\right]}{\partial\varphi^3}\right)\cos(\varphi)\right] \quad (1.15)$$

Здесь $b_\varphi(\varphi,t)$ - безразмерное магнитное поле у основания конвективной зоны Солнца, нормированное на $B_0 \sim \eta_T/R_\odot \sqrt{4\pi\rho_{R_\odot-H_\odot}/\mu}$ и зависящее только от времени и широты. Заметим, что в (1.15) мы для простоты отбросили члены $\mu\tan(\varphi)\left(2f(\varphi)/\cos^2(\varphi)-\partial^2 f(\varphi)/\partial\varphi^2\right)$, обращающиеся в ноль на солнечном экваторе.

### 3. ОЦЕНКИ ВЕЛИЧИНЫ ТИЛЬТА В СОЛНЕЧНОМ ЦИКЛЕ

В данном разделе мы проведем оценки величины тильта в солнечном цикле для реалистичного магнитного поля, полученного в ранее разработанной модели нелинейного самосогласованного динамо, основанной на балансе спиральностей, которая особенно удачно описывает в деталях процесс образования солнечных пятен за последние пять циклов активности, начиная с 20-го солнечного цикла (Safiullin et al., 2018).

Мы считаем наблюдаемые пятна проявлением пороговой неустойчивости, физическая основа которой определяется, в основном, эффектом отрицательного магнитного давления (NEMPI) (Kleeorin et al. 1989; 1990; Rogachevskii and Kleeorin 2007; Brandenburg et al. 2016; Warnecke et al 2016). Эта неустойчивость перераспределяет магнитный поток, созданный механизмом динамо среднего поля. Недавно была построена нелинейная динамо-модель, см. Kleeorin et al. (2003); Zhang et al. (2006); Zhang et al. (2012); Kleeorin et al. 2016), которая принимает во внимание алгебраическую и динамическую части нелинейности альфа-эффекта, опирающаяся на наблюдаемую магнитную спиральность. Эта динамо-модель была откалибрована по длительным рядам наблюдений солнечных пятен, и она позволила рассчитать помесячный прогноз солнечной активности (Safiullin et al. 2018). Мы будем использовать ее для расчета крупномасштабного магнитного поля Солнца.

Используя (1.15), и усреднив по случайной функции $\xi$, получим следующее выражение для среднего тильта:

$$\delta^{(tot)} \approx -\delta_0\left[\sin(\varphi)-C_1\phi_1\cos(\varphi)\right] \quad (2.1)$$

Здесь $\delta_0 = 2\pi(1+k)\tau_D^2/\tau_F T_\odot$, $C_1 = 2a_*\tau_F \ell_0^4 \rho_{R_\odot-H_\odot}/3\tau_0 R_\odot^4 \rho_0 \mu(1-2\varepsilon)$.

Проведем оценки данных величин по порядку величины. Время формирования и всплытия активной области $\tau_F$ обычно составляет от нескольких часов до суток ($\tau_F \approx 10^4-10^5$ с), это время соответствует глубине шкалы плотности при формировании супергранулы $H_\rho \approx 10$ Мм и средней скорости всплытия $u_r \approx 10^2-10^3$ m/s. Характерное время установления движения супергранулы $\tau_D = 2L/u_a$, где $L \approx 20$ Мм, где альфвеновская скорость вблизи поверхности $u_a \approx 1.1\cdot 10^3$ m/s, таким образом $\tau_D \approx 3.6 \cdot 10^4$ с, период вращения Солнца по порядку величины $T_\odot \approx 2.2 \cdot 10^6$ с и $k \approx 1$, что дает следующую оценку по порядку величины для $\delta_0 \approx 0.25 - 0.5$ рад., или в градусной мере $\approx 25°$.

Теперь дадим оценку величинам, входящим в $C_1$. Величина $a_*$ порядка единицы, размеры элементарной ячейки грануляции $\ell_0 \approx 1$ Мм, и соответствующее характерное время, определяемое альфвеновской скоростью, $\tau_0 \approx 900$ с. Таким образом, соотношение времен $\tau_F/\tau_0 \approx 10-100$. Оценим соотношение плотностей у поверхности Солнца и на дне конвективной зоны $\rho_{R_\odot - H_\odot}/\rho_0$ по модели Бэйкера и Тамесвари (1966): вблизи дна $\rho_{R_\odot - H_\odot} \approx 2 \cdot 10^{-1} \frac{\text{g}}{cm^3}$, а на глубине 300 км $\rho_0 \approx 8 \cdot 10^{-7} \frac{\text{g}}{cm^3}$, следовательно отношение плотностей $\rho_{R_\odot - H_\odot}/\rho_0 \approx 2.5 \cdot 10^5$. Отношение масштабов грануляции к радиусу Солнца $\ell_0/R_\odot \approx 1/700$ и, как показано в работе Клиорин, Рузмайкин 1982, $\mu \approx 1/9 \approx 0.11$. Объединяя вышеуказанные оценки, получим $C_1 \approx (0.7-7) \cdot 10^{-6}/(1-2\varepsilon)$. Отсюда, взяв, например, смешанные граничные условия $\varepsilon \approx 0.33$, мы можем для последующих оценок использовать значение $C_1 \approx (0.2-2) \cdot 10^{-5}$.

Наконец, функция $\phi_1$ в рамках «no-r» модели (Соколов и др., 1995; Сафиуллин и др., 2018) имеет вид:

$$\phi_1(\vartheta,t) = \begin{cases} \dfrac{\partial^3\left[B^2(\theta,t)\right]}{\partial \theta^3}, & |B| > B_0 \\ 0, & |B| \leq B_0 \end{cases} \quad (2.2)$$

Следует принимать во внимание, что значения функции $\phi_1$ в рамках модели по факту нормированы на $B_0^2$. Из уравнения (2.1) следует, что полный тильт $\delta^{(tot)}$ в наших предположениях имеет два вклада. Оба они происходят от силы Кориолиса, действующей на поднимающуюся на поверхность активную область, при этом второе слагаемое связано с силой Кориолиса, действующей на глобальную циркуляцию. Коэффициент $\psi$ показывает, какая доля спиральности, существующей на границе конвективной зоны и фотосферы, выносится в корону и хромосферу. Параметры и константы в (2.1), включая пороговое поле $B_0$ и долю спиральности $\psi$, плохо известны в контексте нашей «no-r» модели. Поэтому полезно считать их в нашей модели подгоночными параметрами.

Для расчета вышеупомянутых вкладов в тильт мы используем значения магнитного поля, полученные в модели, разработанной в работе Сафиуллин и др. (2018) и воспроизводящие широтно-временную динамику пятенной активности Солнца, в особенности детально за последние пять циклов (20й-24й), а именно начиная с 1964 г. Однако, при описании образования активных областей, мы будем учитывать не только возникновение пар пятен (для этого в работе Сафиуллин и др. 2018 был предложен режим отсечки по величине магнитного поля $B_0$), но и более широкий класс магнитных проявлений, соответствующий помимо пятен еще и их полутеням, для чего мы снизим иллюстративный порог образования этих структур, например, на 60% от уровня пятен, по сравнению с формулой (2.2), а именно до $0.4\,B_0$, см. рис. 1.

На основании разработанной модели с теми же самыми параметрами мы можем произвести расчет широтно-временной структуры функции вклада силы Кориолиса, действующей на глобальную циркуляцию $\phi_1$, см. рис. 2. В структуре этой функции более отчетливо просматриваются постепенное нарастание со временем нарушения симметрии полушарий относительно экватора, как проявление вековой квадрупольной компоненты глобального магнитного поля.

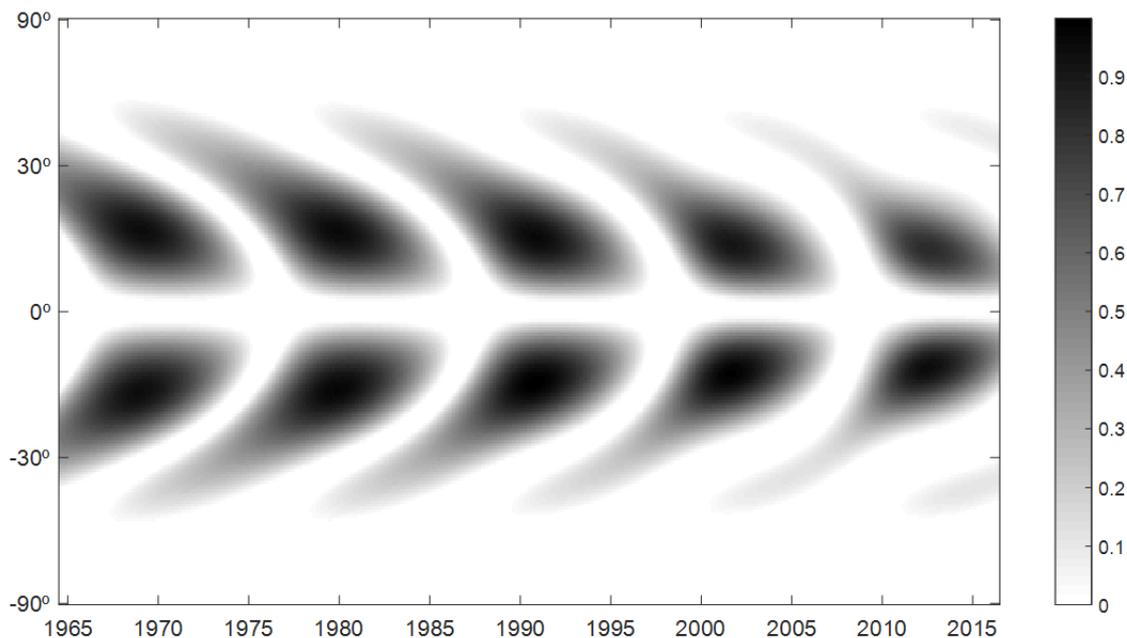

**Рис. 1.** Широтно-временная диаграмма (бабочки) по магнитному полю в модели Safiullin et al. (2018) с величиной порога всплытия $0.4\,B_0$ от порога образования пятен

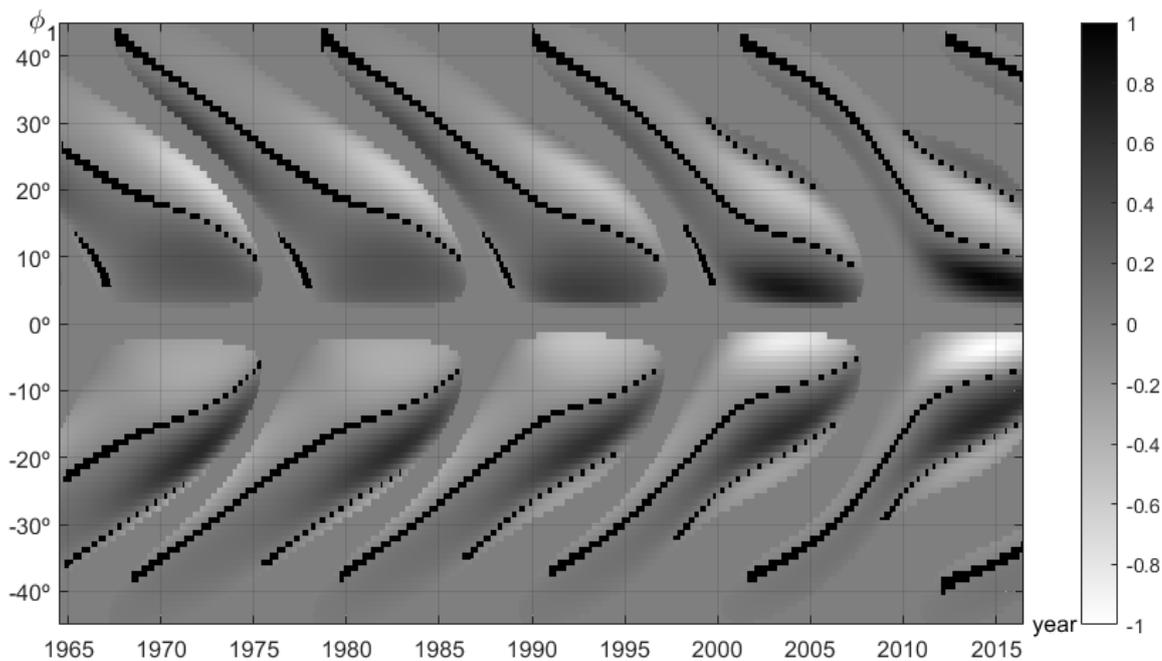

**Рис. 2.** Широтно-временная диаграмма (бабочки) по функции вклада силы Кориолиса, действующей на глобальную циркуляцию $\phi_1$ (нормирована на 1). Черными линиями на бабочках особо выделены линии нуля.

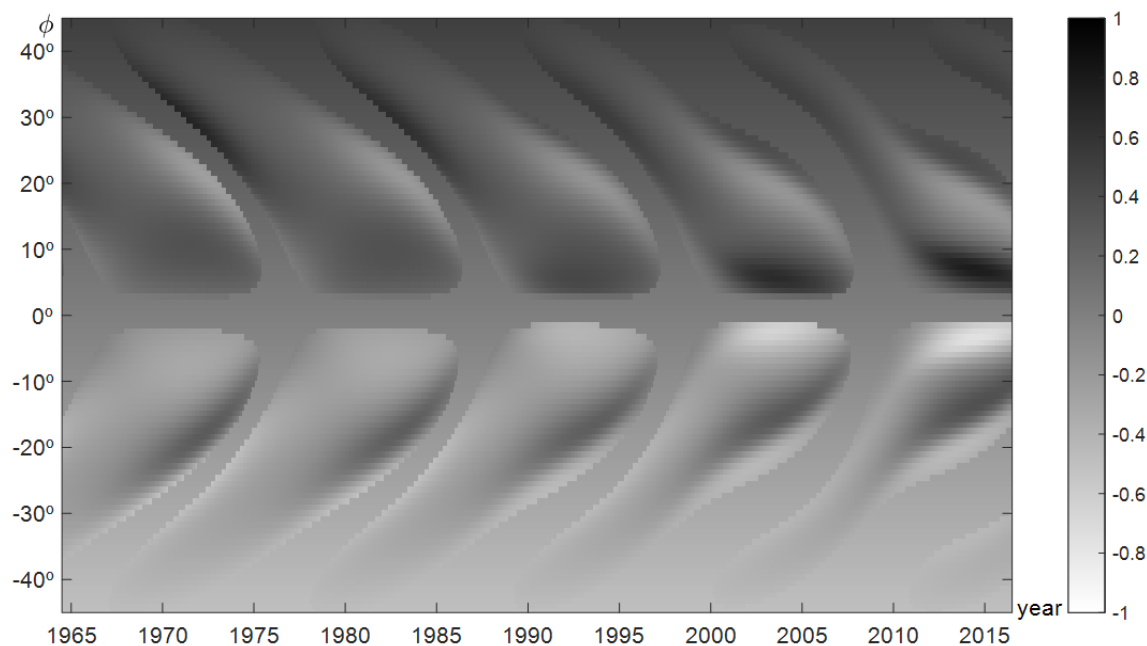

**Рис. 3.** Широтно-временная диаграмма (бабочки) по тильту в долях от величины $\delta_0$, использовано демонстрационное значение константы $C_1$ с величиной порога всплытия $0.4\ B_0$ от порога образования пятен.

Теперь мы можем рассчитать широтно-временную диаграмму бабочек по тильту с демонстрационным значением константы $C_1$, см. рис. 3 (нормировано на 1). При построении Рис. 3 использовалось демонстрационное значение константы $C_1$, которое дает такую нормировку $\phi_1$, что по выражению (2.1) мы получаем значения тильта, похожие на наблюдаемые. При этом надо иметь в виду, что за широтой 30-40 градусов пятен нет и тильт наблюдательно определить невозможно.

Для сравнения результатов этих иллюстративных модельных расчетов с наблюдениями, например, Тлатовой и др. (2015), мы проведем осреднение значений тильта в течение каждого из циклов 20-24. Результаты см. на рис. 4.

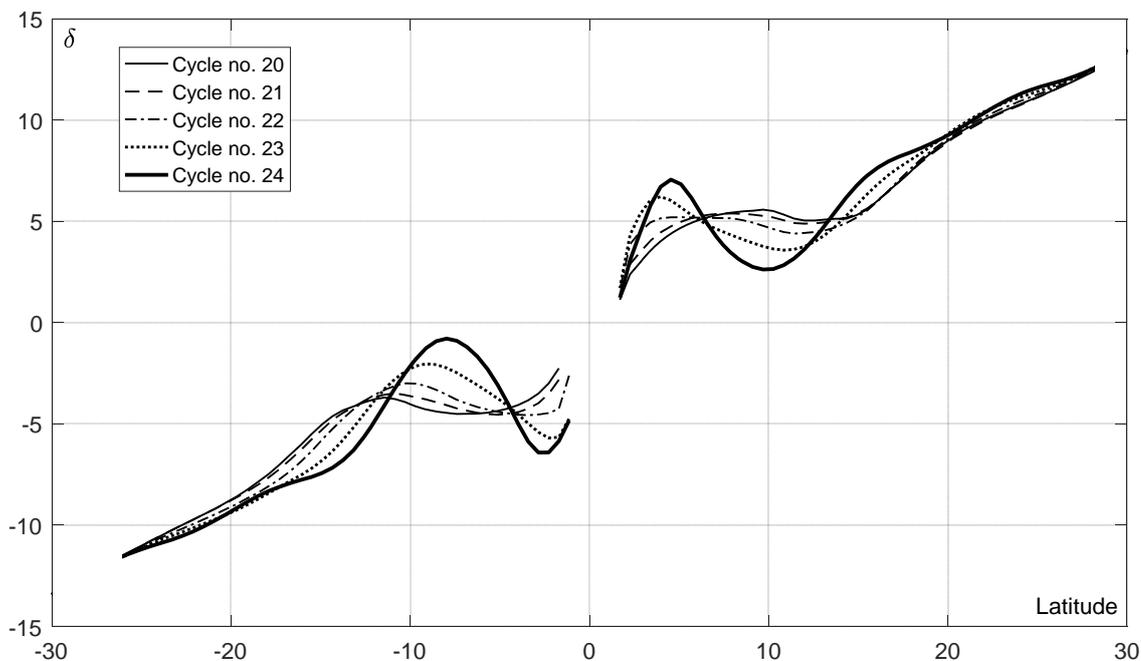

**Рис. 4.** Величины тильта, усредненного по отдельным циклам (20-й - 24й), от широты вблизи экватора.

## 4. ОБСУЖДЕНИЕ

Итак, основная идея нашей работы состоит в том, что, хотя тилт-угол в первом приближении объясняется действием сил Кориолиса (анти-симметричных по широте), значительный вклад в тильт вносится за счет усредненного действия магнитных сил Ампера. Этот вклад имеет как симметричную по широте, так и анти-симметричную части, и он зависит от глобального магнитного поля на Солнце, которое меняется в течение полного магнитного 22-летнего солнечного цикла. При смене полярности в солнечном динамо каждые 11 лет меняется и эта симметричная по широте часть. Это приводит к тому, что вблизи экватора средний тильт может быть отличным от нуля, в четных циклах активности он как правило выше, чем в нечентных. Как показывают рисунки 3 и 4, значения тильта могут даже незначительно менять знак на отдельных фазах солнечного цикла и в определеном диапазоне широт, при этом по усреднении не меняя в среднем общих закономерностей закона Джоя.

    Из приведенных иллюстраций (см. рисунки 3 и 4) видно, что значения тильта существенно варьируются вблизи экватора. Данный эффект находится в соответствии с показанной в работе Тлатовой и др. (2015) вариацией среднего значения тильт-углов вблизи экватора от цикла к циклу. Изменение параметра С1 может даже иногда изменить значение осредненного тильта за весь цикл на определенном диапазоне широт. Однако, наша иллюстрация показывает, что при сделанном выборе параметра С1 можно избежать таких явлений. Данный результат согласуется, например, с работами Тлатова и др. (2014) и (2016).

## 5. ЗАКЛЮЧЕНИЕ

В данной работе мы пытались дать теоретическое объяснение особенностям в поведения тильта, используя результаты модели нелинейного «no-r» динамо, элементарные соображения о закручивании полей активных областей силой Кориолиса и результатов расчета магнитного поля по модели динамо, основанного на законе сохранения магнитной спиральности.

Для простоты, а также понимания физического механизма образования тильта, мы не учитывали наличие внутренней скрученности магнитного поля всплывающей активной области и связанным с ней законом сохранения магнитной спиральности, что будет сделано в последующих работах.

Наша модель даже в рамках самых общих предположений о действии силы Кориолиса на регулярную и случайную компоненты скорости как неоднородного вращательного, так и меридионального движения вещества и взаимодействия с магнитным полем, позволила получить реалистичные оценки величины тильта, его широтно-временную зависимость и средние значения в окрестности экватора. Важно отметить, что в результате рассмотрения даже такой простейшей модели мы получили основные закономерности особенностей поведения и широтно-временной структуры тильта вблизи солнечного экватора. Показано, что вблизи самого экватора могут существовать области заметного отклонения тильта от закона Джоя, которые, однако, в реалистичном диапазоне параметров модели при усреднении не приводят к его полному отрицанию, но заметны на достаточно полном статистически богатом материале современных исследований (см. Тлатов и др., 2016; Тлатова и др. 2015, 2018).

Для понимания деталей поведения тильта по широте и фазе солнечного цикла потребуются дальнейшие исследования, которые буду представлены в последующих работах. Полученные результаты могут представлять интерес не только для описания свойств закрутки активных областей (тильта), но и возможно позволят оценить поток магнитной спиральности в хромосферу и корону и, тем самым, энергетический потенциал вспышечной активности Солнца.

Kleeorin N., Kuzanyan K., Moss D., Rogachevskii I., Sokoloff D. and Zhang H., Magnetic helicity evolution during the solar activity cycle: Observations and dynamo theory". *Astron. Astrophys.*, 2003, **409**, 1097-1105.

Zhang H., Sokoloff D., Rogachevskii I., Moss D., Lamburt V., Kuzanyan K. and Kleeorin N., The radial distribution of magnetic helicity in the solar convective zone: Observations and dynamo theory. *Month. Not. Roy. Astron. Soc.*, 2006, **365**, 276-286.

Zhang H., Moss D., Kleeorin N., Kuzanyan K., Rogachevskii I. , Sokoloff D., Y. Gao and H. Xu, Current helicity of active regions as a tracer of large-scale solar magnetic helicity. *Astrophys. J.*, 2012, **751**, 47.

Kleeorin Ya., Safiullin N., Kleeorin N., Porshnev S., Rogachevskii I. and Sokoloff D., The dynamics of Wolf numbers based on nonlinear dynamos with magnetic helicity: Comparisons with observations. *Month. Not. Roy. Astron. Soc.*, 2016, **460**, 3960-3967.

Kleeorin N., Rogachevskii I. and Ruzmaikin A. Negative magnetic pressure as a trigger of a large-scale magnetic instability in the solar convective zone. *Sov. Astron. Lett.*, 1989, **15**, 274-277.

Rogachevskii and I.Kleeorin N. Shear-current effect in a turbulent convection with a large-scale shear. *Phys. Rev. E*, 2007, **75**, Issue 4, id. 046305

Warnecke J., Losada I. R., Brandenburg A., Kleeorin N. and Rogachevskii I., Bipolar region formation in stratified two-layer turbulence. *Astron. Astrophys.*, 2016, **589**, A125.

Клиорин Н.И. и Рузмайкин А.А. Динамика средней турбулентной спиральности в магнитном поле. *Магнитная гидродинамика*, 1982, № **2**, 17-24.

Sokoloff, D.D., Fioc, M., Nesme-Ribes, E. "Asymptotic properties of dynamo waves"// Соколов Д. Д., Фьок, М., Нем-Риб, Э.; Магнитная Гидродинамика 31, No. 1, 19-38, 1995

Sokoloff D., Bao S.D., Kleeorin N., Kuzanyan K., Moss D., Rogachevskii I., Tomin D., Zhang H., The distribution of current helicity at the solar surface at the beginning of the solar Cycle". *Astron. Nachr.*, 2006, **327**, 876-883.

Baker N. , Tamesvary S. Tables of Convective Stellar Envelop Models. New York: NASA, 1966.